\documentclass{jetpl}
\twocolumn
\lat

\usepackage{epsfig}
\usepackage{graphicx,url,color}
\graphicspath{{./Figures/}}

\usepackage{multirow}

\newcommand{\beq}{\begin{equation}}
\newcommand{\eeq}{\end{equation}}
\newcommand{\bea}{\begin{eqnarray}}
\newcommand{\eea}{\end{eqnarray}}

\newcommand{\eq}{\begin{equation}}
\newcommand{\en}{\end{equation}}
\newcommand{\eqa}{\begin{eqnarray}}
\newcommand{\ena}{\end{eqnarray}}

\title{Decomposition of the static potential in $SU(3)$  gluodynamics}
\rtitle{Decomposition\ldots}

\sodtitle{Decomposition of the static potential in $SU(3)$  gluodynamics}

\author{V.\,G.\,Bornyakov (http://orcid.org/0000-0002-5395-7258)$^{a}$\/\thanks{e-mail: vitaly.bornyakov@ihep.ru},
I.\,E.\,Kudrov$^a$}

\rauthor{V.\,G.\,Bornyakov V.G., I.\,E.\,Kudrov I.E.}

\sodauthor{Bornyakov, Kudrov}

\address{$^a$NRC ``Kurchatov Institute'' - IHEP, 142281 Protvino, Russia}

\begin{document}
\abstract{After fixing the Maximal Abelian gauge in $SU(3)$ lattice gluodynamics
we decompose the nonabelian gauge field
into the Abelian field created by Abelian monopoles and the modified nonabelian field with monopoles
removed. We then calculate respective static potentials in the fundamental
representation and show that the sum of these potentials approximates
the nonabelian static potential with good precision at all distances considered.
Comparison with other ways of decomposition is made.
}



\maketitle


\section{Introduction}
\label{section1}

We study numerically the lattice $SU(3)$
gluodynamics in the Maximal Abelian gauge (MAG)
and consider decomposition of the lattice gauge
field $U_\mu(x) \in SU(3)$
\begin{equation}
 U_\mu(x) =  U_\mu^{mod}(x) U_\mu^{mon}(x)
\label{eq:decomposition}
\end{equation}
where $U_\mu^{mon}(x)$ is the component of the gauge field due to
Abelian monopoles (to be defined later)
and $U_\mu^{mod}(x)$ is respectively the monopoleless
component which we call a modified gauge field.
By modification we understand removal of the Abelian monopoles.

This kind of decomposition was studied before in $SU(2)$ gluodynamics in
 \cite{Bornyakov:2005hf}. It was shown that while the monopole component $U_\mu^{mon}(x)$ is reproducing the linear
part of the static potential, the monopoleless component $U_\mu^{mod}(x)$ produces purely
Coulomb potential and their sum provides a good approximation of the original unaltered static
potential at all distances:
\begin{equation}
 V(R) \approx  V_{mod}(R) + V_{mon}(R)
\label{eq:poten_decomp}
\end{equation}
Recently, in \cite{Bornyakov:2021enf} it was shown that this approximation becomes better when the
lattice spacing is decreased leaving a possibility that  relation (\ref{eq:poten_decomp})
becomes exact in the continuum limit. It was also shown that (\ref{eq:poten_decomp})
is satisfied in $SU(2)$ QCD as well. In the present work we extend the study of the decomposition
(\ref{eq:decomposition}) to the more realistic case - $SU(3)$ gluodynamics.

It is well known \cite{suzuki1,suzuki2,Bali:1996dm,bm,Sakumichi:2014xpa} that
after performing the Abelian projection in the MAG
\cite{Kronfeld:1987ri,thooft2}, the Abelian
string tension calculated from the Abelian static potential
is very close to the nonabelian string tension. This observation, confirmed
in gluodynamics and in QCD, supports the concept of the Abelian dominance
(for a review see e.g. \cite{review}).
It was further discovered \cite{suzuki3,stack,Bali:1996dm} that the so called monopole static
potential also has string tension close to the nonabelian one.
These observations are in agreement with
conjecture that monopole degrees of
freedom are responsible for confinement \cite{thooft}.

The interesting question is what is the
role of the other, i.e. monopoleless
degrees of freedom. The results obtained
in SU(2) gluodynamics suggest that
they are responsible for the Coulomb part of the static potential both at small and
large distances. This suggests that while at small distances  $U_\mu^{mod}(x)$ gives perturbative contribution into the static potential, it provides nonperturbative contribution at large distances.


It is worth to note that the gauge covariant decomposition was introduced in Refs.~\cite{Cho:1979nv} and \cite{Duan:1979} and developed
further in Refs.~\cite{Faddeev,Shabanov:1999xy,Kondo:2005eq}, see for review \cite{Kondo:2014sta}. The numerical results demonstrating analogues of the
Abelian dominance and the monopole dominance within this approach were obtained  in \cite{Kato:2014nka}.
 It would be
interesting to check if the decomposition into the monopole and monopoleless components works in this approach.

The decomposition different from eq.~(\ref{eq:decomposition})
was considered in $SU(3)$ gluodynamics after fixing MAG  \cite{Sakumichi:2014xpa}.
 The usual coset decomposition of the gauge field into the Abelian and the off-diagonal components was used:
\begin{equation}
 U_\mu(x) =  U_\mu^{offd}(x) U_\mu^{Abel}(x)
\label{eq:decomposition2}
\end{equation}
and respective decomposition for the static potential was verified:
\begin{equation}
 V(R) \approx  V_{offd}(R) + V_{Abel}(R).
\label{eq:poten_decomp2}
\end{equation}
We will compare our results for decomposition (\ref{eq:poten_decomp2}) with results
of Ref.~\cite{Sakumichi:2014xpa} in section \ref{sec:poten_decomp}.

The decomposition similar to (\ref{eq:poten_decomp})
for the static potential in the maximal center gauge
\begin{equation}
 V(R) \approx  V_{cent}(R)  + V_{mod,cent}(R)
\label{eq:poten_decomp3}
\end{equation}
corresponding to the decomposition of the gauge field into the center and modified (vortex free) components:
\begin{equation}
 U_\mu(x) =  U_\mu^{mod,cent}(x) U_\mu^{cent}(x) \,.
\label{eq:decomposition3}
\end{equation}
was first checked long ago in $SU(2)$ gluodynamics \cite{Bornyakov:2005hf} and was studied  recently in QCD \cite{Biddle:2022zgw}.
 We will comment on these numerical results later in section \ref{sec:poten_decomp}.

\section{Decomposition of the gauge field}

We consider the SU(3) lattice gluodynamics after fixing MAG.
We use the definition of MAG introduced for lattice
$SU(N)$ theory in \cite{Kronfeld:1987vd}  and later specified for the SU(3)
group in \cite{Brandstater:1991sn}. The MAG is fixed by maximizing the
functional
\beq
F = \frac{1}{8\,V}
 \sum_{x,\mu}\left[ |U^{(11)}_\mu(x)|^2 +|U^{(22)}_\mu(x)|^2
            +|U^{(33)}_\mu(x)|^2 -1 \right]
\eeq
with respect to local gauge transformations $g$ of the lattice gauge field,
\beq
U_\mu(x) \to U^g_\mu(x) = g(x)^\dagger U_\mu(x) g(x+\hat \mu)\,.
\eeq
To fix MAG, the simulated annealing algorithm with three random gauge copies was used.
This algorithm was first used to fix MAG in the
$SU(2)$ case \cite{Bali:1996dm} and then extended to the $SU(3)$ group in
\cite{Bornyakov:2001qw}. The details of  the implementation of the simulated
annealing algorithm in the case of $SU(3)$ gauge group can be found in
\cite{DIK:2003alb}.
For the gauge fixing functional $F$ we obtained the average value $<F>=0.73388(1)$ to be compared with $<F>=0.7322(2)$  quoted in \cite{Sakumichi:2014xpa,Ohata:2020dgn}. The larger is the value of the maximized functional the better is the gauge fixing. The difference in $<F>$ is due to the Gribov copies effects and implies that there might be a substantial difference between our results and results of Ref.~\cite{Sakumichi:2014xpa} for gauge dependent quantities like Abelian string tension as discussed in detail in \cite{DIK:2003alb}.

The Abelian projection means coset decomposition  (\ref{eq:decomposition2})
of the nonabelian lattice gauge field $U_\mu(x) \in SU(3)$  into the Abelian field $U^{Abel}_\mu(x) \in U(1) \times U(1)$ and
the coset field $U^{offd}_\mu(x) \in SU(3)/U(1) \times U(1)$.
The Abelian field $U^{Abel}_\mu(x)$ is determined as
\beq
\label{uabel}
U^{Abel}_\mu(x) = \mbox{diag} \left(u^{(1)}_\mu(x),u^{(2)}_\mu(x),u^{(3)}_\mu(x)
\right)\,,
\eeq
where
\beq
\label{ulink}
u^{(a)}_{\mu}(x)=e^{i \theta^{(a)}_{\mu}(x)}
\eeq
with
\beq
\label{tlink}
\theta^{(a)}_{\mu}(x) = \arg~(U_\mu(x))_a-\frac{1}{3} \sum_{b=1}^3
\arg(U_\mu(x))_b\,\big|_{\,{\rm mod}\ 2\pi}
\eeq
such that
\beq
\theta^{(a)}_{\mu}(x) \in [-\frac{4}{3}\pi, \frac{4}{3}\pi]\,.
\eeq
This definition of Abelian projection $u_\mu(x)$ maximizes the expression
$|\mbox{Tr} \left( U_\mu^\dagger(x) u_\mu(x) \right) |^2$
\cite{DIK:2004xwj}.
The Abelian gauge fields  can in turn be decomposed into
monopole (singular) and photon (regular) parts:
\beq
\theta_{\mu}^{(a)}(x)=\theta^{(a)\,mon}_{\mu}(x)+\theta^{(a)\,ph}_{\mu}(x)\,,
\label{monph}
\eeq
The monopole part is defined by \cite{Smit:1989vg}:
\beq
\theta^{(a)\,mon}_{\mu}(x)= 2 \pi \sum_{y} {D(x-y)\partial^-_{\alpha}
m_{\alpha\mu}^{(a)}(y)}\,, \label{montheta}
\eeq
where integers  $m_{\mu\nu}^{(a)}(x)$ denote the singular part of the
Abelian plaquettes (Dirac plaquettes),  $\partial^-_{\alpha}$ is the backward lattice derivative, and $D(x)$ denotes the lattice Coulomb propagator. Then $U^{mon}_\mu(x)$
introduced in  (\ref{eq:decomposition}) is defined as
\beq
\label{umon}
U^{mon}_\mu(x) = \mbox{diag} \left(e^{i\theta^{(1)\,mon}_\mu(x)},e^{i\theta^{(2)\,mon}_\mu(x)},e^{i\theta^{(3)\,mon}_\mu(x)} \right)\,.
\eeq

\section{The static potential decomposition}
\label{sec:poten_decomp}
\begin{figure}[htb]
\hspace*{0mm}
\includegraphics[width=8.5cm,angle=0]{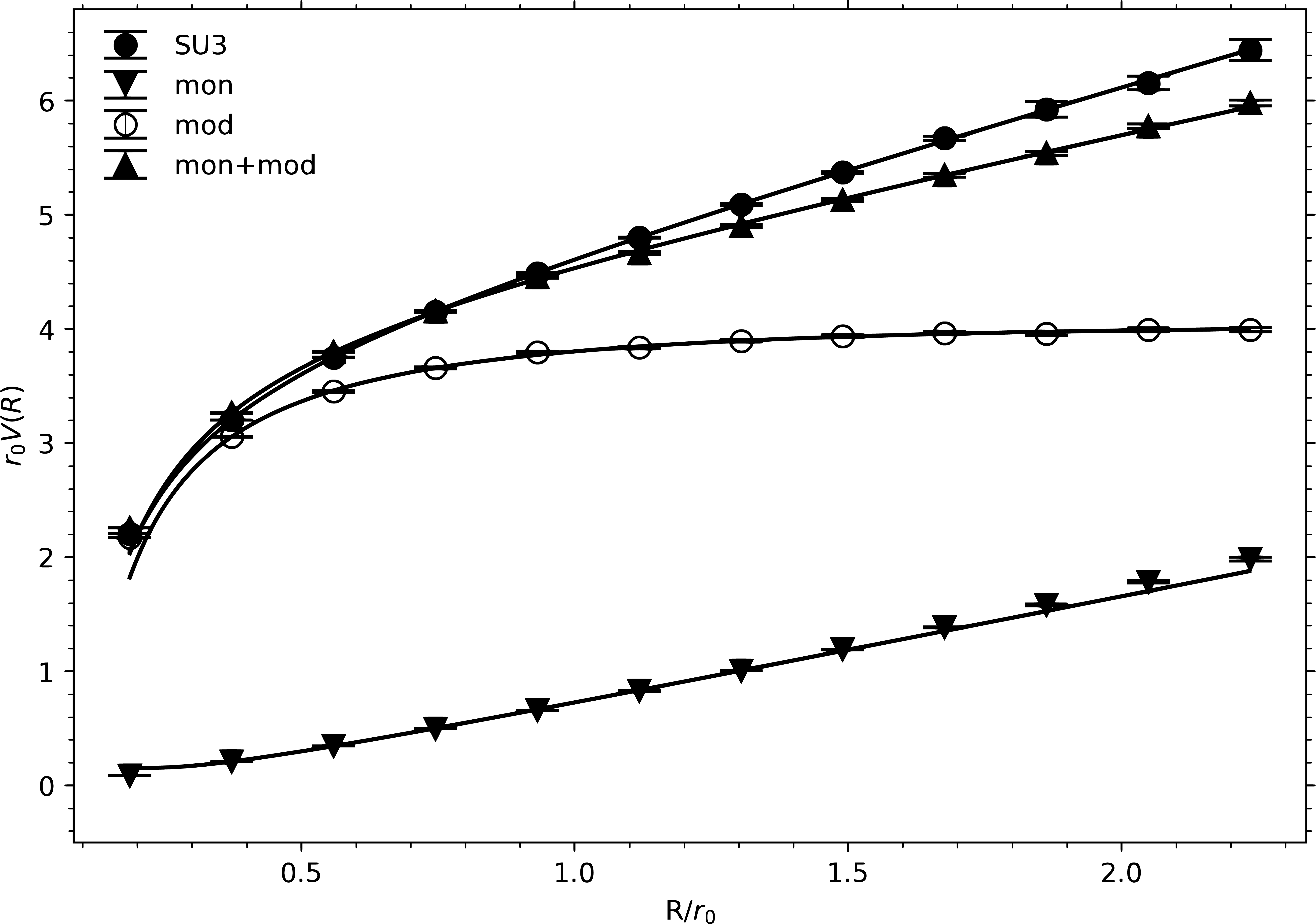}
\caption{Fig~1. Comparison of the nonabelian potential $V(R)$ (filled circles) with the
sum  $V_{mod}(R)+V_{mon}(R)$ (filled triangles) for $\beta=6.0$.
$V_{mod}(r)$ (empty circles) and $V_{mon}(r)$ (filled inverted triangles) are also depicted. The solid curves  show the fits  to  the Cornell potential.}
\label{potentials1}
\end{figure}

We calculated $R \times T$ rectangular Wilson loops $W(R,T)$, $W_{mon}(R,T)$ and $W_{mod}(R,T)$
using lattice gauge fields $U_\mu(x)$,  $U_\mu^{mon}(x)$, $ U_\mu^{mod}(x)$ introduced above. To extract respective static potentials $V(R)$, $V_{mon}(R)$ and $V_{mod}(R)$
 the APE smearing \cite{APE:1987ehd} has been employed. Computations were done with the Wilson lattice action at
$\beta = 6.0$ on $24^4$ lattices using $5000)$ statistically independent configurations.
The lattice spacing at this value of the bare coupling constant is
defined by $a/r_0 = 0.186(4)$
\cite{Necco:2001xg}, where $r_0=0.5$fm is called Sommer parameter.
In Fig.~\ref{potentials1} our results are presented. One can see that similar to the $SU(2)$ gluodynamics \cite{Bornyakov:2021enf} the monopole potential $V_{mon}(R)$ is almost perfectly linear with small curvature at small distances while the modified field potential $V_{mod}(R)$ is well described by the Coulomb potential. It is also seen that
the relation (\ref{eq:poten_decomp}) is valid at all distances with the most essential
discrepancy of about $10 \% $  at large distances. It is clear that this discrepancy is mostly due to
a rather low slope of $V_{mon}(R)$. As we have already mentioned it was found in $SU(2)$ gluodynamics that with decreasing of the lattice spacing agreement in relation  (\ref{eq:poten_decomp}) improves
substantially. This should be checked in SU(3) gluodynamics in the future.

\begin{table}[hbtp]
\begin{center}
\begin{tabular}{|cccc|}  \hline
potential        &  $\sigma  r_0^2$   & $\alpha $    &  $r_0 V_0$      \\
\hline
$V(R)$ & 1.34(2)  &-0.34(1)  & 3.61(2)   \\
$V_{mon}(R)$ &   0.99(1) & 0.09(1) & -0.36(2)  \\
$V_{mod}(R)$
&   0  & -0.42(1) & 4.22(1) \\
$V_{mon}(R) + V_{mod}(R)$ &   0.94(1) & -0.39(1)& 3.98(2)\\
\hline
\end{tabular}
\end{center}
\caption{Table~1. Parameters of the potentials obtained by fits to function
$V_0 - \alpha/R + \sigma R$.}
\end{table}

\begin{figure}[htb]
\hspace*{-5mm}
\includegraphics[width=8.5cm,angle=0]{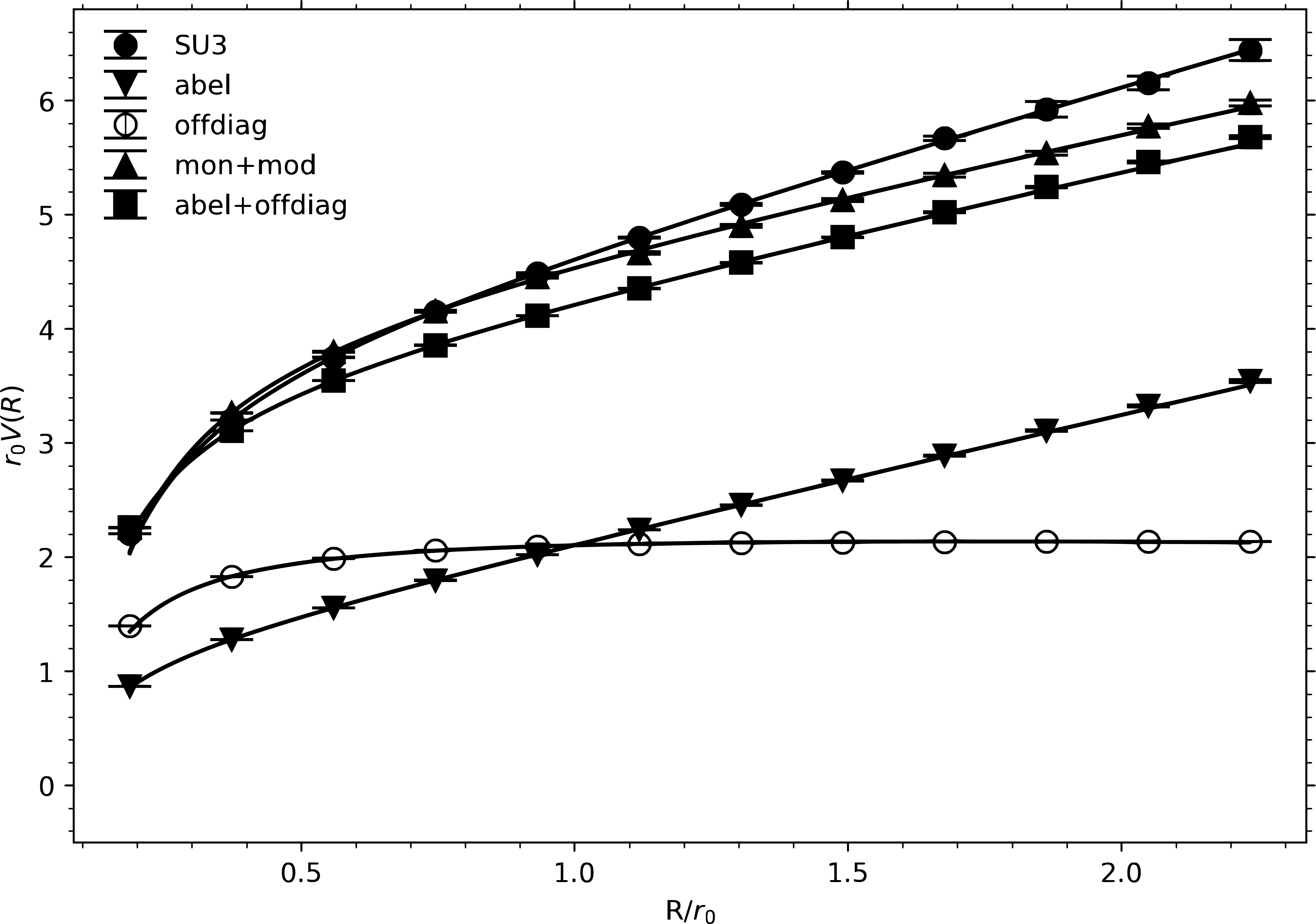}
\caption{Fig~2. Comparison of the nonabelian potential $V(R)$ (filled circles) with the
sum  $V_{mod}(R)+V_{mon}(R)$ (filled inverted triangles) and the sum
$V_{Abel}(R)+V_{offd}(R)$ (filled squares).  $V_{offd}(r)$ (filled triangles) and $V_{Abel}(r)$ (empty circles) are also depicted. The solid curves
show the fits  to  the Cornell potential.}
\label{potentials2}
\end{figure}

Next we come to our results for decomposition (\ref{eq:poten_decomp2}) considered in
Ref.~\cite{Sakumichi:2014xpa}. These results are presented in Fig.~\ref{potentials2}.
In this figure, we compare the original potential $V(R)$  with the decompositions  (\ref{eq:poten_decomp}) and (\ref{eq:poten_decomp2}). One can see that the decomposition
(\ref{eq:poten_decomp}) clearly works better. Both decompositions should be checked in the
continuum limit. In any case, our results clearly contradict the conclusion
made in Ref.~\cite{Sakumichi:2014xpa} that the relation (\ref{eq:poten_decomp2})
is satisfied very nicely already at $\beta=6.0$. Comparing our results presented here
with results we obtained on $16^4$ lattices we found that finite volume effects are small.
We should note that in \cite{Sakumichi:2014xpa} slightly different procedure of the Abelian projection was used but as it was claimed in Ref.~\cite{Stack:2002ysv} the differences between these two procedures of Abelian projection are negligible.
Thus our understanding is that the reason for
such discrepancy with results of  Ref.~\cite{Sakumichi:2014xpa} is the difference in the gauge fixing quality. It was shown in the past that the Gribov copy effects for
gauge non-invariant quantities might be quite substantial \cite{DIK:2003alb,Stack:2002ysv}.

Next, we wish to make a remark about the decomposition (\ref{eq:poten_decomp3}).
This decomposition was first studied in \cite{Bornyakov:2005hf} using the Direct Central gauge in $SU(2)$ gluodynamics.
It was concluded that this decomposition holds with substantially less precision than
the decomposition (\ref{eq:poten_decomp}), first of all, due to the low string tension
provided by the center vortex component. To cure the problem of low string tension
obtained after the center projection the new approach to the definition of the center
gauge was formulated and successfully applied to $SU(2)$   gluodynamics in
\cite{Golubich:2020gea}.
Recently this decomposition was studied in the lattice QCD with light quarks \cite{Biddle:2022zgw}. It was found that in this theory the center projected string tension
is in a very good agreement with the physical string tension. On the other hand, the modified component $U_\mu^{mod,cent}(x)$ produces the static potential which is not
compatible with the Coulomb potential. Thus, the decomposition fails to work at small distances.

\section{Conclusions}
\label{sec:conclusions}

There are a few suggestions for the decomposition of the gauge field into components describing (mostly)
either infrared or ultra-violate physics. These include the gauge covariant Cho decomposition  \cite{Cho:1979nv,Duan:1979,Faddeev,Shabanov:1999xy,Kondo:2005eq}, two decompositions in the MAG, eqs.~(\ref{eq:decomposition}) and (\ref{eq:decomposition2}) and one decomposition in the maximal center gauge (\ref{eq:decomposition3}).
In this work we extended
our study of the decomposition in MAG into the monopole
and the modified (monopoleless) components  in $SU(2)$ gluodynamics  and $SU(2)$ QCD
 \cite{Bornyakov:2021enf} to the case of $SU(3)$ gluodynamics.
We presented our results for one lattice spacing to demonstrate that the decomposition works quite well.
Our results obtained in \cite{Bornyakov:2021enf}  for $SU(2)$ gluodynamics give hope that the decomposition (\ref{eq:decomposition}) will work
even better when the lattice spacing will be decreased.

  Our results for another MAG decomposition,
(\ref{eq:decomposition2}), contradict to results from Ref.~\cite{Sakumichi:2014xpa}
and also indicate that the decomposition (\ref{eq:decomposition}) is superior. There is another reason to consider the
decomposition (\ref{eq:decomposition}) better motivated physically than the decomposition (\ref{eq:decomposition2}). The decomposition (\ref{eq:decomposition}) separates out the  monopole component $U^{mon}_\mu(x)$ which is responsible for the linear part of the static potential as well as for the chiral symmetry breaking. The modified (monopoleless) component $U^{mod}_\mu(x)$ produces purely Coulomb potential
which is in agreement with the original Coulomb part both at small and large distances.
At the same time in the decomposition (\ref{eq:decomposition2}) the Coulomb part is distributed in an unnatural way between two components: Abelian and off-diagonal.

It is clear that the study of both decompositions at varying lattice spacing is necessary to
understand their fate in the continuum limit. This is the subject of our future work.

\bibliographystyle{apsrev}

\begin{thebibliography}{99}

\bibitem{Bornyakov:2005hf}
V.~G.~Bornyakov, M.~I.~Polikarpov, G.~Schierholz, et al,
Nucl. Phys. B Proc. Suppl. \textbf{153}, 25-32 (2006)
[arXiv:hep-lat/0512003 [hep-lat]].

\bibitem{Bornyakov:2021enf}
V.~G.~Bornyakov, I.~Kudrov and R.~N.~Rogalyov,
Phys. Rev. D \textbf{105}, no.5, 054519 (2022)
[arXiv:2101.04196 [hep-lat]].

\bibitem{suzuki1}
T.~Suzuki and I.~Yotsuyanagi, Phys.\ Rev.\ D {\bf 42} (1990) 4257.
\bibitem{suzuki2}
S.~Hioki, S.~Kitahara, S.~Kiura, et al,
Phys.\ Lett.\ B {\bf 272}, 326 (1991)
[Erratum-ibid.\ B {\bf 281}, 416 (1992)].
\bibitem{Bali:1996dm}
G.~S.~Bali, V.~Bornyakov, M.~Muller-Preussker and K.~Schilling,
Phys.\ Rev.\ D {\bf 54} (1996) 2863.
\bibitem{bm}
V.~Bornyakov and M.~Muller-Preussker,
Nucl.\ Phys.\ Proc.\ Suppl.\  {\bf 106}, 646 (2002).
\bibitem{Sakumichi:2014xpa}
N.~Sakumichi and H.~Suganuma,
Phys. Rev. D \textbf{90} (2014) no.11, 111501.
\bibitem{Kronfeld:1987ri}
A.~S.~Kronfeld, M.~L.~Laursen, G.~Schierholz and U.~J.~Wiese,
Phys.\ Lett.\ B {\bf 198}, 516 (1987).
\bibitem{thooft2}
G.~'t Hooft, Nucl.\ Phys.\ B {\bf 190} (1981) 455.
\bibitem{review}
M.~N.~Chernodub and M.~I.~Polikarpov,
in "Confinement, Duality and Non-perturbative Aspects of QCD", p.387, Plenum Press, 1998, hep-th/9710205;
R.~W.~Haymaker,
Phys.~Rept.~{\bf 315} (1999) 153;
J.~Greensite,
Prog. Part. Nucl. Phys. \textbf{51} (2003), 1
\bibitem{suzuki3}
H.~Shiba and T.~Suzuki,
Phys.\ Lett.\ B {\bf 333}, 461 (1994).
\bibitem{stack}
J.~D.~Stack, S.~D.~Neiman and R.~J.~Wensley,
Phys.\ Rev.\ D {\bf 50}, 3399 (1994).
\bibitem{thooft}
G.~'t~Hooft, in {\it High Energy Physics}, ed. A. Zichichi,
EPS International Conference, Palermo (1975);
S.~Mandelstam, Phys.\ Rept.  {\bf 23}, 245 (1976).

\bibitem{Cho:1979nv}
Y.~M.~Cho,
Phys. Rev. D \textbf{21} (1980), 1080
doi:10.1103/PhysRevD.21.1080
\bibitem{Duan:1979}
Y.S. Duan and M.L. Ge, Sinica Sci., 11, 1072–1081 (1979).
\bibitem{Faddeev}
L. Faddeev and A.J. Niemi,
Phys. Rev. Lett. 82, 1624–1627 (1999);
L.D. Faddeev and A.J. Niemi,
Nucl. Phys. B776, 38-65 (2007).
\bibitem{Shabanov:1999xy}
S.V. Shabanov,
Phys. Lett. B458, 322–330 (1999);
S.V. Shabanov,
Phys. Lett. B463, 263–272 (1999).
\bibitem{Kondo:2005eq}
K.~I.~Kondo, T.~Murakami and T.~Shinohara,
Prog. Theor. Phys. \textbf{115} (2006), 201-216;
S.~Kato, K.~I.~Kondo, T.~Murakami, A.~Shibata, T.~Shinohara and S.~Ito,
Phys. Lett. B \textbf{632} (2006), 326-332.
\bibitem{Kondo:2014sta}
K.~I.~Kondo, S.~Kato, A.~Shibata and T.~Shinohara,
Phys. Rept. \textbf{579} (2015), 1-226.
\bibitem{Kato:2014nka}
S.~Kato, K.~I.~Kondo and A.~Shibata,
Phys. Rev. D \textbf{91} (2015) no.3, 034506.
\bibitem{Biddle:2022zgw}
J.~C.~Biddle, W.~Kamleh and D.~B.~Leinweber,
Phys. Rev. D \textbf{106} (2022) no.5, 054505
doi:10.1103/PhysRevD.106.054505
[arXiv:2206.00844 [hep-lat]].
\bibitem{Kronfeld:1987vd}
A.~S.~Kronfeld, G.~Schierholz and U.~J.~Wiese,
Nucl. Phys. B \textbf{293} (1987), 461-478.
\bibitem{Brandstater:1991sn}
F.~Brandstater, U.~J.~Wiese and G.~Schierholz,
Phys. Lett. B \textbf{272} (1991), 319-325.
\bibitem{Bornyakov:2001qw}
V.~Bornyakov, G.~Schierholz and T.~Streuer,
Nucl. Phys. B Proc. Suppl. \textbf{106} (2002), 676-678
doi:10.1016/S0920-5632(01)01813-8
[arXiv:hep-lat/0111018 [hep-lat]].
\bibitem{DIK:2003alb}
V.~G.~Bornyakov \textit{et al.} [DIK],
Phys. Rev. D \textbf{70} (2004), 074511
doi:10.1103/PhysRevD.70.074511
[arXiv:hep-lat/0310011 [hep-lat]].
\bibitem{Ohata:2020dgn}
H.~Ohata and H.~Suganuma,
Phys. Rev. D \textbf{102} (2020) no.1, 014512.
\bibitem{DIK:2004xwj}
V.~G.~Bornyakov \textit{et al.} [DIK],
Phys. Rev. D \textbf{71} (2005), 114504
doi:10.1103/PhysRevD.71.114504
[arXiv:hep-lat/0401014 [hep-lat]].
\bibitem{Smit:1989vg}
J.~Smit and A.~van der Sijs,
Nucl. Phys. B \textbf{355} (1991), 603-648
\bibitem{APE:1987ehd}
M.~Albanese \textit{et al.} [APE],
Phys. Lett. B \textbf{192} (1987), 163-169
\bibitem{Necco:2001xg}
S.~Necco and R.~Sommer,
Nucl. Phys. B \textbf{622} (2002), 328-346.

\bibitem{Stack:2002ysv}
J.~D.~Stack, W.~W.~Tucker and R.~J.~Wensley,
Nucl. Phys. B \textbf{639} (2002), 203-222
doi:10.1016/S0550-3213(02)00537-0
[arXiv:hep-lat/0110196 [hep-lat]].
\bibitem{Golubich:2020gea}
R.~Golubich and M.~Faber,
Particles \textbf{3} (2020) no.2, 444-455.




\end{thebibliography}

\end{document}